\documentstyle[12pt,fleqn,espcrc1,epsfig]{article}
\begin{document}
\newcommand{\ttbs}{\char'134}
\newcommand{\AmS}{{\protect\the\textfont2
  A\kern-.1667em\lower.5ex\hbox{M}\kern-.125emS}}
\hyphenation{author another created financial paper re-commend-ed}


\title{Hard Rescattering in QCD and High Energy Two-Body Photodisintegration 
of the Deuteron}

\author{Leonid L. Frankfurt\address{ School of Physics and Astronomy,
Tel Aviv University, Tel Aviv 79978, Israel \\
$^{b}$Department of Physics, University of Washington, Box 351560,
Seattle WA 98195, USA \\
$^{c}$Department of Physics, Florida International University, 
Miami, FL 33199, USA \\
$^{d}$Yerevan Physics Institute, Yerevan 375036, Armenia\\
$^{e}$Department of Physics, Pennsylvania State University,
  University Park, 16802, USA\\
$^{f}$Deutsches Electron Synchroton, DESY, Germany }
Gerald A. Miller$^{b}$,
  Misak M. Sargsian$^{c,d}$ and Mark I. Strikman$^{e,f}$}

\maketitle

\begin{abstract}

Photon absorption by a quark in one nucleon followed by its high momentum
transfer interaction with a quark in the other may produce two nucleons
with high relative momentum. We sum the relevant quark rescattering
diagrams, to show that the scattering amplitude depends on a convolution
between the large angle $pn$ scattering amplitude, the hard photon-quark
interaction vertex and the low-momentum deuteron wave function. The computed
cross sections are in reasonable agreement with the data.
\end{abstract}

\vspace{-0.2cm}
\section{Introduction}

\vspace{-0.4cm}

The experiments on high energy two-body photodisintegration of the
deuteron\cite{E89012,NE17} set a new stage in high energy
(E$_\gamma \ge$ 1 GeV) nuclear physics. The conventional mesonic picture 
of nuclear interactions failed to describe the qualitative features of  
these measurements. Thus these experiments are unique in testing the 
implications of quantum chromodynamics QCD in nuclear
reactions~\cite{BCh,Holt}.

One of the first predictions for $\gamma d \rightarrow pn$ reactions 
within QCD was that according to the quark counting rule: 
${d\sigma/dt\sim s^{-11}}$.
This prediction was based on the hypothesis that the Fock state 
with the minimal number of partonic constituents will dominate in 
two-body large angle hard collisions\cite{hex}. Although successful in 
describing energy dependences of number of hard processes, this
hypothesis does not allow to make calculation of the absolute 
values of the cross sections. 
Especially for reactions involving baryons, the calculations within 
perturbative QCD underestimate  the measured cross sections by orders 
of magnitude see e.g.\cite{Isgur_Smith}. 
This may be the indication that in the accessible range of energies 
bulk of the interaction is in the domain of the nonperturbative
QCD\cite{Isgur_Smith,Rady}. On the other hand even if we fully realize 
the role of the nonperturbative interactions the theoretical methods of 
calculations are very restricted.

Here we investigate the effects in which the absorption of the photon
by a quark of one  nucleon, followed by  a high-momentum transfer
(hard) rescattering with a quark from the second nucleon, produces the 
final two nucleon state of large relative momenta.
We demonstrate that the structure of hard interaction for this 
rescattering mechanism is similar to that of hard NN scattering.
Therefore the sum of the multitude of diagrams with incalculable 
nonperturbative part of the interaction is expressed through 
the experimentally measured amplitude of hard $np$ scattering.

Another important feature of discussed mechanism is that its
dominant contribution comes from the low relative momentum
($<$ 300 MeV/c) of two nucleons. Therefore for the deuteron 
wave function one can use conventional wave functions calculated 
using the realistic nucleon-nucleon potentials. 


\section{Kinematic Requirements}
\vspace{-0.4cm}

The use of the partonic picture requires that 
the masses of the intermediate hadronic state produced by the 
$\gamma N$ interaction be larger than some minimum mass characterizing
the threshold to reach the continuum. This is known from deep inelastic
scattering\cite{Feynman} to be $W\approx 2.2~GeV$. Here  the mass of
the intermediate (between the photon absorption and quark rescattering)
virtual state is $m_{int}\sim \sqrt{2E_\gamma m_N}$. {}From the condition
$m_{int}\ge W$ one obtains $E_{\gamma}\ge 2.4~GeV$.
Next, the struck quark (Fig.~1) should be energetic enough to reach the
quark of the other nucleon without radiating soft (bremsstrahlung) gluons.
From the regeneration length of the soft gluonic field one obtains the 
condition: $E_{\gamma} \ge r_{N}/R_{regen}^2\sim 2~GeV$.
Finally, to ensure that the quark rescattering is hard enough, one requires 
that: $-t,-u\ge 2~GeV^2$.

\vspace{-1.2cm}
\begin{figure}[htb]
\centering{
\epsfig{angle=0,width=8.0cm,height=2.0cm,file=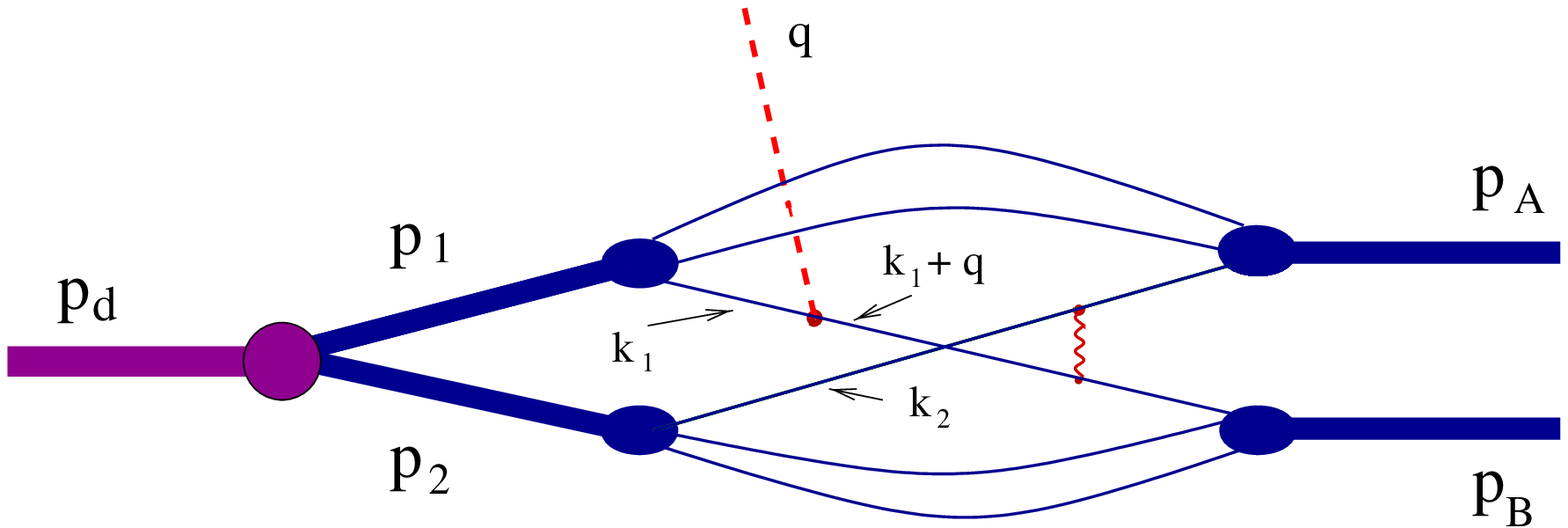}}
\label{Fig.1}
\end{figure}
\vspace{-0.8cm}
\noindent
{Figure.1 \ Quark Rescattering diagram.}

\section{Derivation of Differential Cross Section}
\vspace{-0.4cm}

Our derivation proceeds by evaluating Feynman diagrams such as Fig.~1.
The quark interchange mechanism, in which quarks are exchanged between
nucleons via the exchange of a gluon, is used. All other quark-interactions
are included in the  partonic wave function of the nucleon, $\psi_N$.
We use  a simplified notation in which only the momenta of the
interacting quarks require labeling.
The scattering amplitude $T$ for photo-disintegration of a deuteron
(of four-momentum $p_d$ and  mass $M_d$) into two nucleons of momentum
$p_A$ and $p_B$ is given by:  
\vspace{-0.23cm}
\begin{eqnarray}
&& T = -\sum\limits_{e_q} \int  \left( 
{\psi_N^\dag(x'_2,p_{B\perp},k_{2\perp})\over x'_2}\bar u(p_B-p_2+k_2)\right. 
\left[-igT_c^{F}\gamma^{\nu}\right]
{u(k_1+q)\bar u(k_1+q)\over (k_1+q)^2 - m_q^2 + i\epsilon}\nonumber \\
&&\left[-ie_q\epsilon^{\perp}\cdot\gamma^{\perp}\right]
\left. u(k_1)
{\psi_N(x_1,p_{1\perp},k_{1\perp})\over x_1}\right)\left\{
{\psi_N^\dag(x'_1,p_{A\perp},k_{1\perp})\over x'_1}
\bar u(p_A-p_1+k_1)\left[-igT_c^{F}\gamma_{\mu}\right]\right.\nonumber \\
&&\left . u(k_2) 
{\psi_N(x_2,p_{2\perp},k_{2\perp})\over x_2}\right\}
G^{\mu\nu} {\Psi_{d}(\alpha,p_{\perp})\over 1-\alpha}
{dx_1\over 1-x_1} {d^2k_{1\perp}\over 2(2\pi)^3}
{dx_2\over 1-x_2} {d^2k_{2\perp}\over 2(2\pi)^3}
{d\alpha\over \alpha} {d^2p_\perp\over 2(2\pi)^3},
\label{Ta}
\end{eqnarray}
where $p_1$ and  $p_2$ are the momenta of the nucleons in the deuteron,
with $\alpha \equiv {p_{1+}\over p_{d+}}$, $p_2=p_d-p_1$ and $p_{1\perp}=
-p_{2\perp}\equiv p_\perp$.
Each nucleon consists of one active  quark of momenta $k_1$  and $k_2$: 
$x_i\equiv {k_{i+}\over p_{i+}} = {k_{i+}\over \alpha p_{d+}}$ ($i=1,2$).
$G^{\mu\nu}$ describes the gluon exchange between interchanged quarks.
We use the reference frame where $p_d = (p_{d0},p_{dz},p_{\perp})\equiv
({\sqrt{s'}\over 2}+{M_d^2\over 2\sqrt{s'}},
{\sqrt{s'}\over 2}-{M_d^2\over 2\sqrt{s'}} ,0)$,
with  $s = (q+p_d)^2$, $s'\equiv s-M_D^2,$ and the photon four-momentum is
$q = ({\sqrt{s'}\over 2}, -{\sqrt{s'}\over 2},0)$.

To proceed  we analyze  the denominator of the knocked-out quark propagator,
when recoil quark-gluon system with mass $m_R$ is on mass shell.
We are concerned with  momenta such that $p^2_\perp\ll m^2_N\ll s'$ 
and $\alpha \sim {1\over 2}$ so we neglect terms of order
$p_\perp^2,m^2_N/s'\ll 1$ to obtain:
\begin{equation} 
(k_1+q)^2-m_q^2 + i\epsilon \approx x_1s^\prime(\alpha-\alpha_c + i\epsilon),
\ \mbox{with} \ \alpha_c \equiv {x_1 m_R^2+k_{1\perp}^2\over 
(1-x_1)x_1\tilde s},
\label{alphac}
\end{equation}
where  $\tilde s \equiv s'(1+{M_d^2\over s'})$.
Next we calculate the photon-quark hard scattering vertex 
and integrate over the $\alpha$ using 
only the pole contribution in Eq.(\ref{alphac}). 
Note that the dominant 
contribution arises from the soft component of the deuteron when
$\alpha_c={1\over 2}$, which according to eq.(\ref{alphac}) 
requires $k_{1\perp}^2\approx {(1-x_1)x_1\tilde s\over 2}$.

Summing over the struck quark contributions from
photon scattering off neutron and proton  one can express
the scattering amplitudes through the $pn$ hard scattering amplitude  
within QIM-$A_{pn}^{QIM}(s,l^2)$ as follows:
\begin{eqnarray}
T \approx   
{ie (\epsilon^++\epsilon^{-})(e_u + e_d) \over  2 \sqrt{s'}}
\int f({l^2\over s})A_{pn}^{QIM}(s,l^2) \Psi_{d} ({1\over 2},p_{\perp})
{d^2p_\perp\over (2\pi)^2}.
\label{Tc}
\end{eqnarray}
where $\epsilon^\pm={1\over 2}(\epsilon_x\pm i \epsilon_y)$ and 
$e_u$ and $e_d$ are the electric charges of $u$ and $d$ quarks.
The factor $f(l^2/s)$ accounts for the difference between  the hard
propagators  in our process and those occurring in wide angle $pn$
scattering. Within the  Feynman mechanism\cite{Feynman},  the interacting 
quark carries the whole momentum of the nucleon  ($x_{1}\rightarrow 1$), 
thus $f(l^2/s)=1$.
Within the minimal Fock state approximation, $f(l^2/s)$ 
is the scaling function of the $\theta_{cm}$ only with 
$f(\theta_{cm}=90^0)\approx 1$\cite{gdpn}.

We compute the differential cross section averaging $|T|^2$ over the spins
of initial photon and deuteron and summing over the spins of the  final
nucleons. Then we use the observation that the quark interchange topologies
are the dominant for fixed $\theta_{cm}=90^0$ high
momentum transfer (non strange) baryon-baryon 
scattering. Thus in the region of $\theta_{cm}\approx 90^0$ we
replace $A_{pn}^{QIN}$ by the experimental data - $A_{pn}^{Exp}$ and 
obtain\cite{gdpn}:
\begin{equation}
{d\sigma^{\gamma d\rightarrow pn }\over dt}  =  {4\alpha\over 9}\pi^4\cdot
{1\over s'} C({t_N\over s}){d\sigma^{pn\rightarrow pn}(s,t_N)\over dt}
\times \left| \int\Psi_d^{NR}(p_z=0,p_\perp)\sqrt{m_N}
{d^2p_\perp\over (2\pi)^2}\right|^2,
\label{difcrsb}
\end{equation}
where $t_{N} = (p_B-p_d/2)^2$.
Eq. (\ref{difcrsb}) shows that the ${d\sigma^{\gamma d\rightarrow pn}\over dt}$
depends on the soft component of the deuteron wave function, the  measured
high momentum transfer $pn\rightarrow pn$ cross section and the scaling factor
$C({t_N\over s})\approx f^2(t_N/s)\approx 1$ at $\theta_{cm}\sim 90^0$
(and  slowly varying as a function of $\theta_{cm}$) and the 
additional factor coming from the $\gamma-q$ interaction.

\vspace{-0.2cm}
\section{Comparison with the Data}
\vspace{-0.4cm}

In the numerical calculations we take $C({t_N\over s})=1$ and use
$\Psi^{NR}_{d}$ calculated with Paris potential.
Our calculation produces band because of the accuracy of experimental 
data on ${d\sigma^{pn\rightarrow pn}\over dt}$ 
and our interpolation to the required $(s,t)$ bins. 
Figure 2 shows that calculations are in agreement with the measured
differential cross sections. Moreover the agreement improves for larger
$\theta_{cm}$ which confirms our expectation that $C(t_N/s)\approx 1$
at $\theta_{cm}=90^0$.  
The deviations from the calculation based on the approximation 
$C(\theta_{cm})=1$ seem to be consistent with $C$
being a function of $\theta_{cm}$ only. 
In particular the whole set of available data can  be described (Fig.3) by  
taking $C(t_N/s)={-2t_N/s'\over 1+2t_N/s'}\approx {-t/s'\over 1+t/s'}$
including even the data at $\theta_{cm}=36^0$
where ($-t\le 2~GeV^2$) and the hard interaction mechanism can not be applied.
This may indicate that connection between $NN$ and $\gamma d\to pn$ dynamics
extends to a transitional region of $t$.

\vspace{-0.8cm}
\begin{figure}[htb]
\hspace{-7.5cm} 
\centering{
\epsfig{file=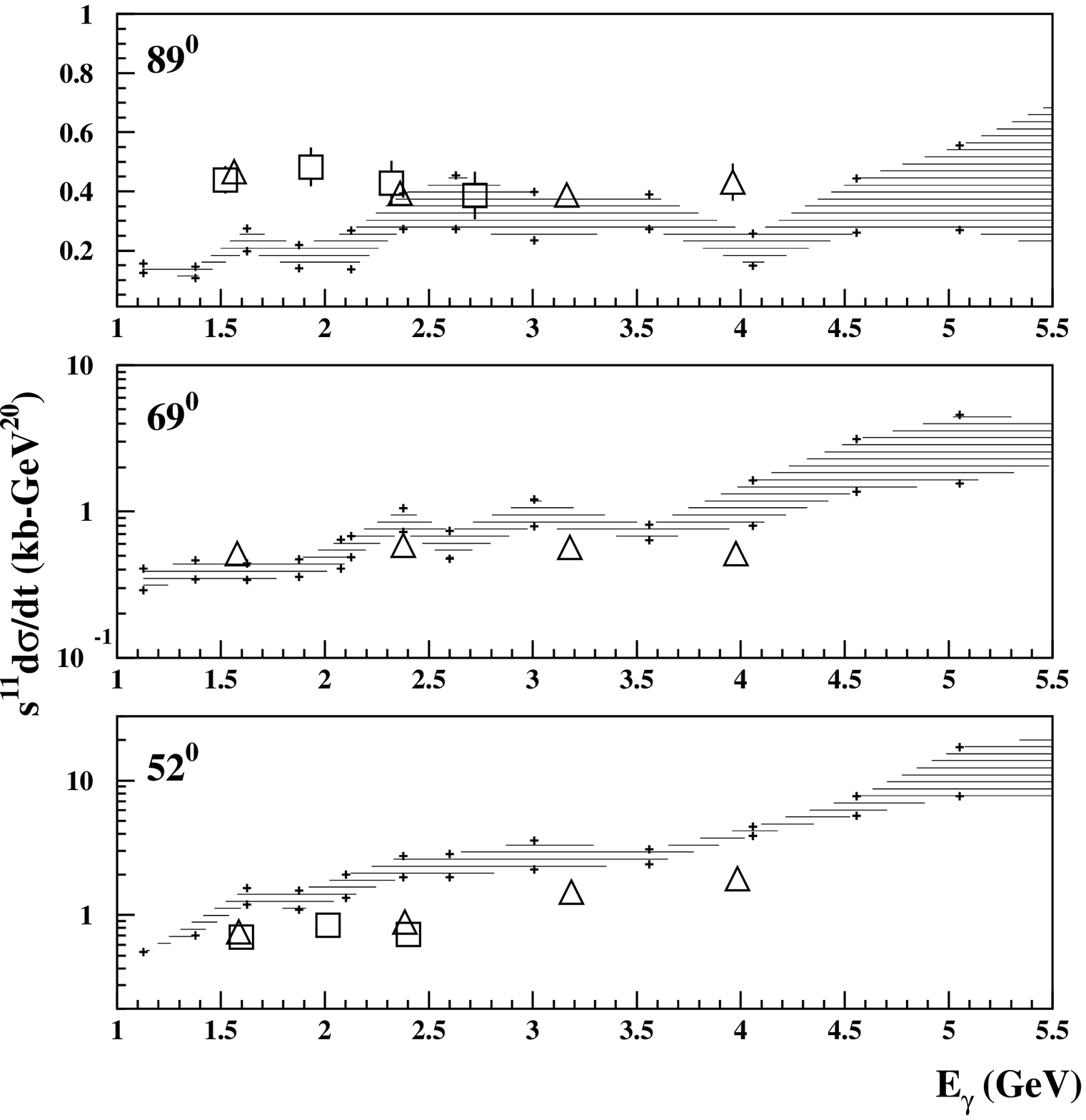,width=7.9cm,height=5.0cm}}  
\label{Fig.2}
\end{figure}
\vspace{-1.0cm}
\noindent
{Figure 2. \ The ${d\sigma/dt}s^{11}$ as a function of $E_{\gamma}$.  \\
Data are from [1] (triangles)and [3] (squares).\\

\begin{figure}[htb]
\vspace{-7.4cm}
\hspace{7.6cm}  
\centering{
\epsfig{file=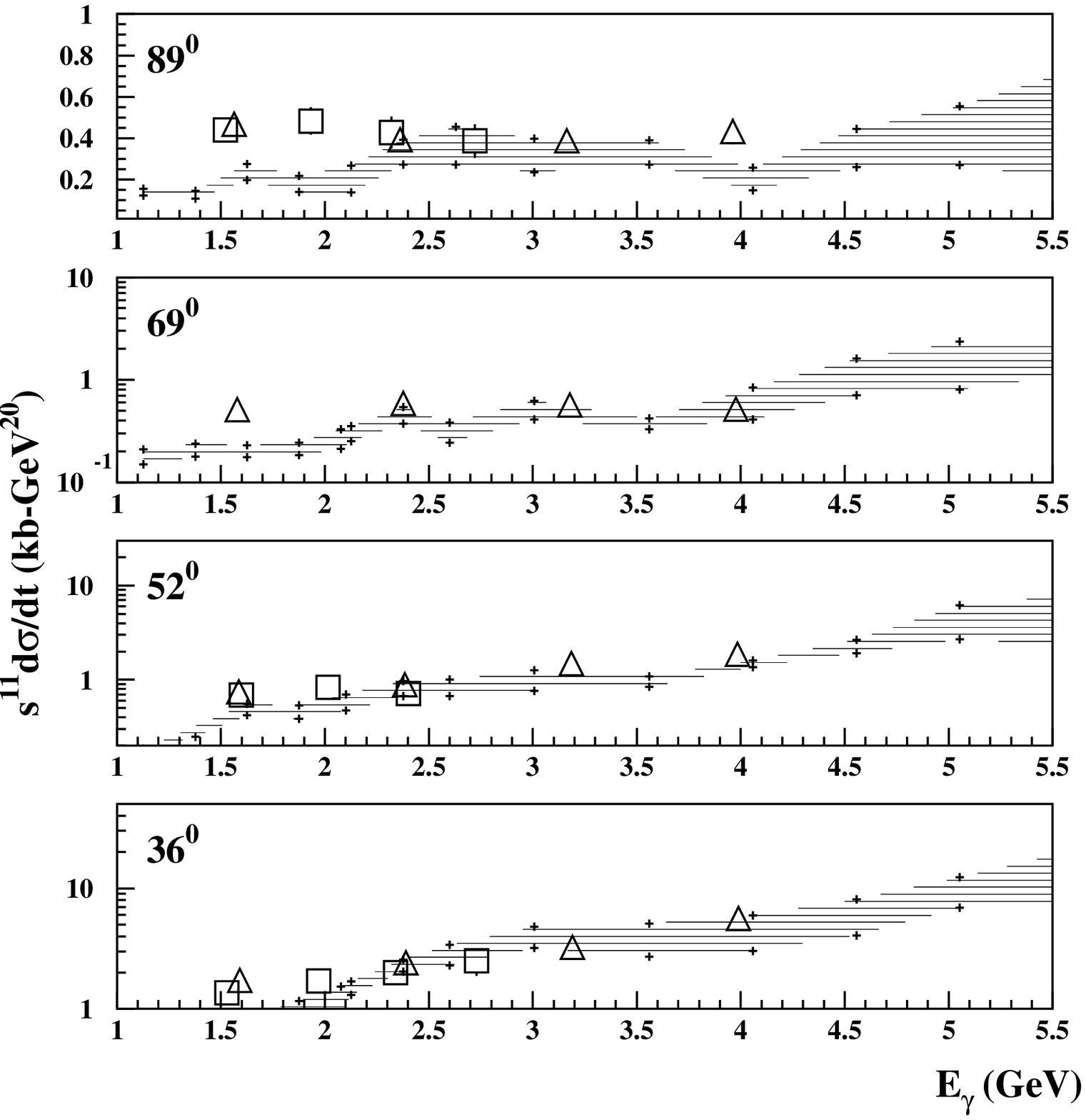,width=7.9cm,height=5.0cm}}

\hspace{8cm} {Figure.3 \  
The same as in Figure.2, with\\   
\hspace{8.2cm} $C(t_N/s)={-2t_N/s'\over 1+2t_N/s'}$. }
\label{Fig.3}
\end{figure}


\medskip

\vspace{-1.6cm}
\section{Summary and Outlook}
\vspace{-0.4cm}

The agreement with the data verifies our underlying
hypothesis  that the size  of the photoproduction reaction is determined by
the physics of high-momentum transfer contained in the hard scattering NN
amplitude. The short-distance aspects of the deuteron wave function are
not important. This  hypothesis, if confirmed by additional studies, 
may suggest the existence of new type of quark-hadron ``duality'', where the
sum of the ``infinite'' number of quark interactions could be replaced by the
hard amplitude of $NN$ interaction.


More data,  especially with a two proton final state
(i.e. $\gamma+^3He\rightarrow pp$ (high $\ p_t$) + n $(p_t\approx 0$)), and a
more detailed angular distribution would definitely allow 
to verify this hypothesis.
The polarization measurement also will be crucial especially at the same $s$
where anomalies observed in hard $pp$ scattering.
Another important extension would be similar experiments using virtual photons.

The present calculations could be extended also to
the reactions with a different composition of final high $p_t$ hadrons,
which could  allow the study the mechanism of the rescattering different from
the quark interchange.


\vspace{-0.4cm}
\begin{thebibliography}{99}
\vspace{-0.4cm}
\bibitem{E89012}C.~Bochna {\em et al.}, Phys. Rev. Lett. {\bf 81}, 4576 (1998).
\bibitem{NE17}J.E.~Belz {\em et al.}, Phys. Rev. Lett. {\bf 74}, 646 (1995).  
\bibitem{BCh}S.J.~Brodsky and B.T.~Chertok, Phys. Rev. Lett. {\bf 37}, 269
        (1976). 
\bibitem{Holt}R.J.~Holt, Phys  Rev. {\bf C41}, 2400 (1990).
\bibitem{hex}S.J.~Brodsky and G.R.~Farrar, Phys. Rev. Lett. {\bf 31}, 1153;  
        V.~Matveev, R.M.~Muradyan and A.N.~Tavkhelidze, Lett. Nuovo
        Cimento {\bf 7}, 719 (1973).
\bibitem{Isgur_Smith}N.~Isgur and C.H.~Llewellyn Smith, Phys. Rev. Lett.
        {\bf 52}, (1984) 1080.
\bibitem{Rady}A.~Radyushkin, Acta Phys. Pol. {\bf B15}, 403 (1984).
\bibitem{Feynman} R.~Feynman, {\em Photon Hadron Interactions}, W.A. Benjamin
        Inc., 1972.
\bibitem{gdpn}L.L.~Frankfurt, G.A.~Miller, M.M.~Sargsian and M.I.~Strikman,
        hep-ph/9904222.

\end{thebibliography}
\end{document}